\documentclass[transmag, a4paper]{IEEEtran}

\usepackage{amsmath}   % From the American Mathematical Society
\usepackage{amssymb}
\usepackage{amsfonts}  % A popular package that provides many helpful commands
\usepackage{graphicx}  % Required for graphics, photos, etc.
\usepackage{url}       % Written by Donald Arseneau
\usepackage{cite}
\usepackage{citesort}   % nicer looking citations
\usepackage{balance}
\usepackage{subfig}   
\usepackage{xr}
\usepackage{multirow}
\usepackage{standalone}
\usepackage{pgf}
\usepackage{pgfplots}
\usepackage{verbatim}
\usepackage{floatrow}
\usepackage{booktabs}
\usepackage[most]{tcolorbox}
\usepackage[europeanresistors]{circuitikz}
\definecolor{darkgreen}{RGB}{29, 125, 74} 
\newfloatcommand{capbtabbox}{table}[][\FBwidth]
\usepackage{tikz}
\usetikzlibrary{arrows.meta,positioning, shapes, patterns, calc, through, arrows,quotes, angles}
\tikzstyle{vertex}=[circle, draw, inner sep=1pt, minimum size=6pt]
\pgfplotsset{compat=1.18}
\newcommand{\wye}{\mathbin{\tikz[x=1ex,y=1ex]{\draw[line width=.1ex] (0,0)--(30:1)--++(-30:1) (30:1)--++(0,1);}}}
\newcommand{\norm}[1]{\left\lVert#1\right\rVert}
\newcommand*\circled[1]{\tikz[baseline=(char.base)]{
    \node[shape=circle,draw,inner sep=0pt] (char) {#1};}}

\IEEEoverridecommandlockouts 

\begin{document}

\renewcommand{\baselinestretch}{1.0}

\title{AITEE - Agentic Tutor for Electrical Engineering}

\author{Christopher~Knievel, Alexander~Bernhardt, Christian Bernhardt
  \thanks{C.~Knievel, A.~Bernhardt and C.~Bernhardt are with the
    Department of Electrical Engineering and Information Technology,
    HTWG Hochschule Konstanz, University of Applied Sciences, Germany
    (email: \{cknievel,abernhard,cbernhard\}@htwg-konstanz.de).} }

\IEEEtitleabstractindextext{\begin{abstract}

    Intelligent tutoring systems combined with large language models
    offer a promising approach to address students' diverse needs and
    promote self-efficacious learning. While large language models
    possess good foundational knowledge of electrical engineering
    basics, they remain insufficiently capable of addressing specific
    questions about electrical circuits. In this paper, we present
    AITEE, an agent-based tutoring system for electrical engineering
    designed to accompany students throughout their learning process,
    offer individualized support, and promote self-directed learning.
    AITEE supports both hand-drawn and digital circuits through an
    adapted circuit reconstruction process, enabling natural
    interaction with students. Our novel graph-based similarity
    measure identifies relevant context from lecture materials through
    a retrieval augmented generation approach, while parallel Spice
    simulation further enhances accuracy in applying solution
    methodologies.  The system implements a Socratic dialogue to
    foster learner autonomy through guided questioning. Experimental
    evaluations demonstrate that AITEE significantly outperforms
    baseline approaches in domain-specific knowledge application, with
    even medium-sized LLM models showing acceptable performance. Our
    results highlight the potential of agentic tutors to deliver
    scalable, personalized, and effective learning environments for
    electrical engineering education.
  \end{abstract}

}
% make the title area
\maketitle

% \IEEEpeerreviewmaketitle

%%%%%%%%%%%%%%%%%%%%%%%%%
% SECTION: INTRODUCTION %
%%%%%%%%%%%%%%%%%%%%%%%%%
%%%%%%%%%%%%%%%%%%%%%%%%%
\section{Introduction}
\label{sec:Introduction}

\IEEEPARstart{T}{he} field of educational technology has seen
remarkable advancements, with the emergence of transformative tools
such as Learning Management Systems, Massive Open Online Courses, and
Intelligent Tutoring Systems. These technologies have enabled a shift
towards distance learning models, allowing students to learn at their
own pace and providing teachers with the ability to scale up effective
teaching practices~\cite{WOLLNYAreWeThere2021}. However, despite these
innovations, many educational technologies do not substantially change
the traditional role of teachers. Typical teaching activities, such as
providing feedback, motivation, and content adaptation, are still
primarily entrusted to human instructors, leading to the
"teacher-bandwidth problem" where there is a shortage of teaching
staff to provide highly informative and competence-oriented feedback
at large scale~\cite{Wiley2002}.  The advent of ChatGPT, an application based on
state-of-the-art GPT language models for  natural language processing (NLP) model, has further
expanded the potential of Intelligent Tutoring Systems (ITS). Tracing
its origins to the pioneering ELIZA chatbot developed in 1966, the
capabilities of modern chatbots have become increasingly
sophisticated, with the ability to engage in human-like conversations
and provide personalized learning experiences~\cite{Labadze2023}.
Intelligent Tutoring Systems promise to address the limitations of
traditional educational technologies by incorporating computational
models to provide individualized learning, formative feedback, and
personalized learning
paths~\cite{GRAESSERIntelligentTutoringSystems2001}. Chatbots, as a
subtype of dialog systems, have emerged as a particularly promising
approach, with the ability to simulate conversational partners and
provide feedback through natural
language~\cite{WINKLERUnleashingPotentialChatbots2018,WOLLNYAreWeThere2021}.
Despite their potential, deploying chatbots as Intelligent Tutoring Systems
involves several complications. Due to their susceptibility to hallucinations
and limited robustness, unsupervised chatbot  usage may enable students to extract incorrect solutions from the system, which is particularly a problem for weaker
students~\cite{LEHMANNCognitivemetacognitivemotivational2014,MAYNEZFaithfulnessFactualityAbstractive2020,QUIROGAPEREZRediscoveringusechatbots2020,MARINReviewPracticalApplications2021}. Additionally,
there is a risk that students lose their sense of self-efficacy when
solving tasks independently due to excessive support and instead
develop a dependency on the
tutor~\cite{WIGGINSYouThinkYou2017,MARGULIEUXSelfRegulationSelfEfficacyFear2024}.

\begin{figure}[bp]
  \centering \includegraphics[width=0.5\textwidth,
  keepaspectratio]{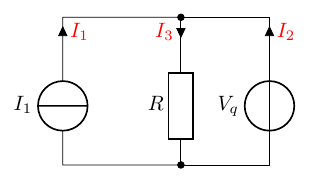}
  \caption{Exemplary electrical circuit with current and voltage
    source as well as an ohmic resistor.}
  \label{fig:circuit_ex}
\end{figure}
\begin{figure*}[tp!]
  \centering \scalebox{0.75}{\input{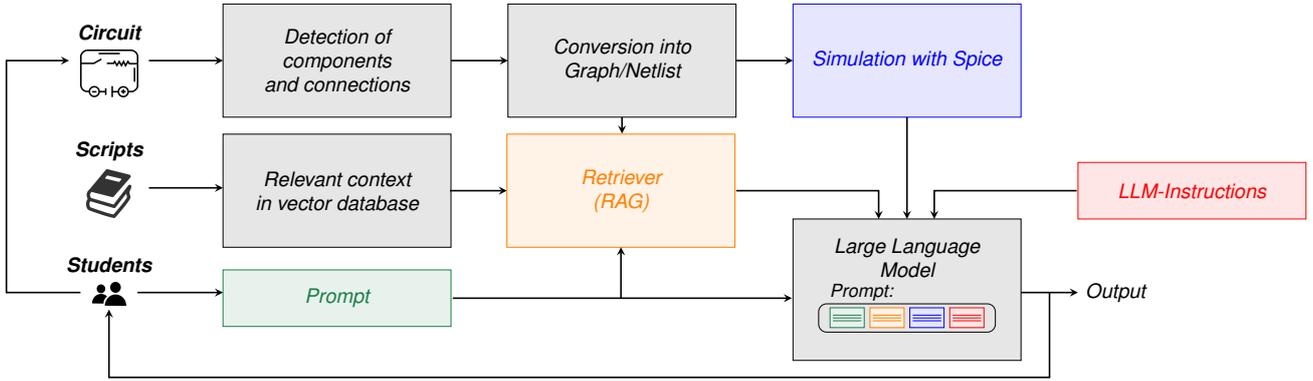}}
  \caption[Componentes of AITEE]{Overview of the required components
    of AITEE.}
  \label{fig:aitee-components}
\end{figure*}
The application of intelligent tutoring systems to electrical
engineering is very
limited~\cite{NEGNEVITSKYApplicationintelligenttutoring1996} and is
restricted to static knowledge representation, lacking dynamic
inference and application of knowledge to solve questions related to
electrical circuits. In this paper, we develop an agentic tutor for
electrical engineering (AITEE), which provides students with an
interactive platform for asking questions about electrical circuits
while ensuring reliability and accuracy of information, leveraging
domain-specific contextual knowledge, and preventing excessive trust
in and dependence on technology. To support students' self-efficacy,
AITEE employs a Socratic dialogue that fosters learner autonomy
through systematic questioning, guiding students toward logical
conclusions~\cite{FAVEROEnhancingCriticalThinking2024,ZHANGSPLSocraticPlayground2024}. Furthermore,
AITEE has to address the typical challenges faced by first-semester
electrical engineering students when analyzing DC circuits, who need
to apply both mathematical foundations, such as linear algebra, as
well as electrical engineering principles, such as Kirchhoff's laws,
to a given circuit. This involves identifying and applying the
solution approaches discussed in the lecture.  An exemplary circuit is
shown in Fig.~\ref{fig:circuit_ex}, with the task of calculating the
current $I_3$ through the ohmic resistance. The challenge for AITEE is
to identify the relevant context within the knowledge base given only
the image of the circuit and the fragmented question: ``How do i
calculate the current $I_3$?'' as input.

We develop a deep learning-based approach to detect the electrical
components and their connections. Different representations of the
query and the circuits were examined for their suitability for
retrieval augmented generation. Due to the poor performance of naive
and advanced RAG methods, we adapted the so-called passage
retrieval~\cite{CHENDenseRetrievalWhat2024} to use a representation of
electrical circuits as indexes, termed index-circuits, and thereby
identify the relevant passages in the script. In order to find the
relevant index{-}circuit for a given query{-}circuit, a similarity
measure between the two circuits must be calculated. For this purpose,
the circuit is transformed into a latent vector representation using a graph-neural network, which captures, among other things, the structure of the circuit. A similarity measure is calculated based on the cosine
distance between the vectors of different circuits. Given the relevant context,
several language models, both open-source as well as closed-source
were evaluated regarding their understanding of electrical circuits
and their ability to correctly solve first-semester electrical
engineering problems. Additionally, the models' robustness against
erroneous information in multi-turn dialogues with students was
investigated.

The remainder of this paper is organized as follows:
Chapter~\ref{sec:system architecture} introduces the architecture of
the agentic system. In chapter.~\ref{sec:repr--simil} the
identification of the electrical circuit as well as the
graph-representation and the subsequent similarity measure are
discussed. Four different large language models (LLMs) are evaluated
in Chapter~\ref{sec:llm-based-tutor} concerning their capabilities of
understanding electrical circuits. Furthermore, the performance of all
four LLMs is evaluated with various prompting and retrieval
strategies. Finally, chapter~\ref{sec:conclusion} concludes the paper.

\section{System Architecture}
\label{sec:system architecture}

Chatbots in education have the potential to increase students'
motivation to learn and strengthen their self-perception and
self-efficacy~\cite{WINKLERUnleashingPotentialChatbots2018}. For an
Intelligent Tutoring System (ITS) in electrical engineering to achieve
these goals, it must be able to understand electrical circuits and
solve tasks by applying the correct methods. However, "AI
hallucinations" - convincingly formulated but factually incorrect
responses - remain an unsolved
problem~\cite{PLEVRISChatbotsPutTest2023}. This is particularly
concerning when students receive these false answers, as they often
lack the ability to verify their correctness. In order to enhance
accessibility and provide seamless learning support, AITEE is designed
to process both digitally created as well as hand-drawn circuit
diagrams. This capability allows students to interact naturally with
the system, whether they are working with computer-generated
schematics or sketching circuits during problem-solving sessions.

AITEE combines several key technologies: circuit image processing to
create netlists (a textual representation of an electrical circuit), a
graph neural network-based similarity measure for context retrieval,
and an LLM supported by Retrieval-Augmented Generation (RAG). Guided by
its system prompt, the tutoring agent engages students in a Socratic
dialogue, promoting active learning and self-efficacy by leading them
towards solutions rather than providing immediate answers. To ensure
accuracy and prevent hallucinations, a SPICE simulation of the circuit
is used to provide precise voltage and current values in case specific
values are given in the task description. These components work
together to create a reliable and effective tutoring system. The
overall architecture of AITEE is shown in
Fig.~\ref{fig:aitee-components}, visualizing the flow of information.
\section{Representation \& Similarity of Electrical Circuits}
\label{sec:repr--simil}

The transformation of hand-drawn circuit diagrams into a
machine-readable format begins with the detection of components and
their interconnections. While research in electrical circuit
recognition is extensive, studies specifically addressing hand-drawn
circuits remain
limited~\cite{REDDYHandDrawnElectricalCircuit2021,BOHARAComputerVisionbased2022,UZAIRElectroNetEnhancedModel2023}. Hand-drawn
circuit recognition presents unique challenges, primarily requiring
robust detection algorithms that can handle inherent imprecisions in
sketches. Notable approaches using YOLO models for component detection  have demonstrated
promising results, achieving AP$_{0.5}$ scores of $98.2\%$ and
$91.6\%$
respectively~\cite{REDDYHandDrawnElectricalCircuit2021,BOHARAComputerVisionbased2022}. Uzair
et al. further refined this approach by developing a two-stage
detector specifically optimized for smaller component
detection~\cite{UZAIRElectroNetEnhancedModel2023}.  The established
method for connection detection in hand-drawn electrical circuits involves a multi-step process: first
removing identified components from the image, then applying Canny
edge detection followed by Hough transformation. The resulting nodes
are then grouped using k-means clustering, with cluster centers
serving as connection endpoints. While this approach has proven
effective for conventional circuits, it faces limitations when applied
to educational contexts.  In educational settings, circuit layouts
often follow specific didactic principles. For instance, star ($\wye$)
or delta ($\Delta$) circuits may intentionally incorporate diagonal
connections or components to emphasize particular circuit
characteristics. These pedagogically motivated layouts present unique
challenges that existing connection recognition methods cannot easily
address. Although the Connected Component Analysis~\cite{HEconnectedcomponentlabelingproblem2017,UZAIRElectroNetEnhancedModel2023}  could be a potential solutions, it is not well-suited for processing hand-drawn circuits due to its susceptibility to the inherent inaccuracies of the given circuits. Therefore, we have developed a novel approach that better
serves these educational requirements.

The following sections describe the technical components of the
circuit analysis system. First, we introduce the netlist as a generic
circuit representation format and the graph neural network for
determining circuit similarities. Next, we present the methods for
object detection and node recognition. The final section details the
calculation of graph embeddings and the similarity measure.

\subsection{Generic Representation}
\label{sec:gener-repr}

\begin{figure}
  \begin{floatrow}
    \ffigbox{ \includegraphics[width=0.5\textwidth,
      keepaspectratio]{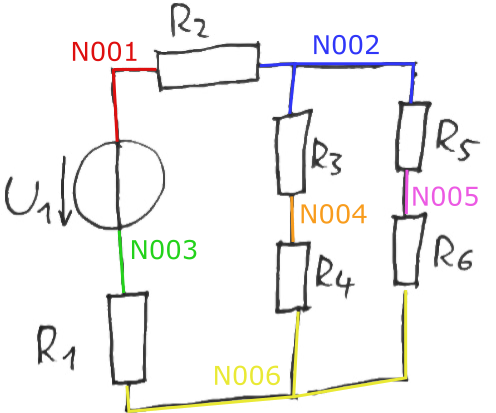} }{\caption{Image of a
        circuit with netlist nodes.}  \label{fig:circuit_color} }
    \capbtabbox{
      \begin{tabular}[h]{ccc}
        R1 & N003 & N006 \\
        R2 & N002 & N001 \\
        R3 & N004 & N002 \\
        R4 & N006 & N004 \\
        R5 & N005 & N002 \\
        R6 & N005 & N005 \\
        U1 & N001 & N003 \\
      \end{tabular}
    }{
      \caption{Netlist of the circuit shown to the
        left.}\label{tab:netlist_ex}
    }
  \end{floatrow}
\end{figure}

\begin{figure}[htbp]
  \centering \begin{circuitikz}[>=latex,
font=\sffamily\slshape, 
thin, 
common/.style={fill=white, font=\sffamily\slshape,
minimum size=0.75cm, inner sep=0pt}
]
\node[draw,circle,minimum size=0.65cm,inner sep=0pt,fill=blue!20!white] (u1) at (0,-0.5) {$U_1$};
\node[draw,circle,minimum size=0.65cm,inner sep=0pt,fill=green!80!black] (r2) at (1,1) {$R_2$};
\node[draw,circle,minimum size=0.65cm,inner sep=0pt,fill=yellow!80!black] (n2) at (3,1) {$N_2$};
\node[draw,circle,minimum size=0.65cm,inner sep=0pt,fill=green!80!black] (r3) at (2,0) {$R_3$};
\node[draw,circle,minimum size=0.65cm,inner sep=0pt,fill=green!80!black] (r5) at (4,0) {$R_5$};
\node[draw,circle,minimum size=0.65cm,inner sep=0pt,fill=green!80!black] (r4) at (2,-1) {$R_4$};
\node[draw,circle,minimum size=0.65cm,inner sep=0pt,fill=green!80!black] (r6) at (4,-1) {$R_6$};
\node[draw,circle,minimum size=0.65cm,inner sep=0pt,fill=yellow!80!black] (n6) at (3,-2) {$N_6$};
\node[draw,circle,minimum size=0.65cm,inner sep=0pt,fill=green!80!black] (r1) at (1,-2) {$R_1$};

\draw (u1) -- (r2) -- (n2) -- (r5) --(r6) -- (n6) -- (r1) -- (u1);
\draw (n2) -- (r3) -- (r4) -- (n6);
\end{circuitikz}
  \caption{Graph representation of the exemplary circuit.}
  \label{fig:graph_circuit_ex}
\end{figure}
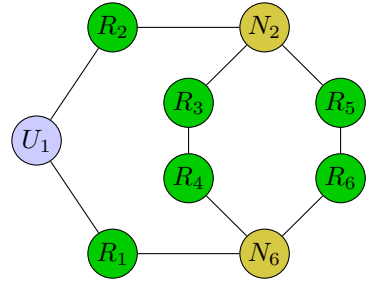

In electrical engineering, the circuit provides the context for a
student's question, with explicit references to specific circuit
elements. To identify relevant solution methods from lecture
materials, AITEE must search for approaches applied to circuits with
similar characteristics, as it cannot be expected that all possible
circuit variations are comprehensively documented.  However, LLMs face
challenges in interpreting graphical representations of electrical
circuits\cite{MESHRAMElectroVizQAHowwell2024}. Netlists, which provide
a textual description of a circuit topology, offer a machine-readable
alternative. The netlist of the circuit shown in
Fig.~\ref{fig:circuit_color} is given as an example in
Table~\ref{tab:netlist_ex}. The netlist of a circuit contains a list
of all components and the corresponding nodes they are connected with,
i.e. in the given example from N001 to N006. It is important to note
that subtle changes in the circuit configuration can significantly
alter the solution strategy. For instance, replacing resistor $R_6$
with a second voltage source $U_2$ requires the use of, for example,
the superposition principle, which is a significant change for a
first-semester student. In the netlist, however, only two characters
are changed. A measure of similarity between two circuits on the basis
of netlists is therefore challenging. Nevertheless, a netlist is used
as input for the supporting SPICE simulation.
A more promising solution compared to the netlist representation is
given by graph neural networks~
\cite{MIRHOSEINIgraphplacementmethodology2021,SAIDCircuitdesigncompletion2023,YAMAKAJICircuit2GraphCircuitsGraph2024}. A
central idea in this paper, is to use the cosine distance between two
feature vectors of a GNN as a measure of similarity between two
electrical circuits. Initially, all components listed in the netlist
were stored as graph nodes. Additionally, connection nodes appearing
more than twice in the netlist were also created as graph nodes,
thereby, enabling the representation of parallel structures within the
graph.  Subsequently, all graph nodes are connected by edges using the
connection nodes from the netlist. The result is a graph that captures
the complete structure of an electrical circuit. Each graph node
stores specific features: node type, number of neighbors, and
centrality, which serve as node embeddings.  The resulting graph of
the circuit in Fig.~\ref{fig:circuit_color} is shown in
Fig.~\ref{fig:graph_circuit_ex}.
For the calculation of a graph similarity, the graph neural network
$\Phi$, parameterized by the weights $\theta$, maps each circuit $c_i$
into an embedding space of $d$
dimensions~\cite{MOHANDeepMetricLearning2023}:
\begin{align}
  \label{eq:1}
  f_i = \Phi\left(c_i;\theta_{\phi}\right)
\end{align}
where $f_i \in \mathbb{R}^d$ is referred to as the feature
representation of the circuit $c_i$. The similarity between two
circuit representations can be calculated by the cosine
similarity~\cite{MOHANDeepMetricLearning2023}:
\begin{align}
  \label{eq:2}
  S(f_i,f_j) = \frac{f_i^{\mathrm{T}} f_j}{\norm{f_i} \cdot \norm{f_j}}.
\end{align}
Cosine similarity provides a measure of vector alignment in space. A
value of $1$ means vectors point in identical directions ($0^\circ$
angle). A value of $0$ indicates perpendicular vectors ($90^\circ$
angle). A value of $-1$ shows vectors pointing in opposite directions
($180^\circ$ angle)~\cite{ALLENAnalogiesExplainedUnderstanding2019}.

For circuit embeddings, this similarity metric captures structural
relationships. Similar circuits have embeddings that point in nearly
the same direction in latent space, with cosine similarity approaching
$1$. As circuits become more dissimilar, their embeddings become
increasingly orthogonal, with cosine similarity nearing $0$.

\subsection{Object Detection \& Node Recognition}
\label{sec:object-detection}

\begin{figure}[htbp]
  \centering \includegraphics[width=0.9\textwidth,
  keepaspectratio]{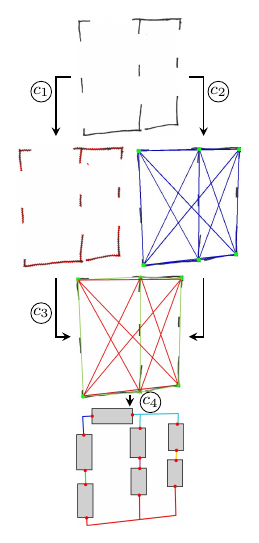}
  \caption{Illustration of the process for recognizing the connection
    nodes in an electrical circuit.}
  \label{fig:edge-detect-flow}
\end{figure}

Similarly
to~\cite{MIRHOSEINIgraphplacementmethodology2021,SAIDCircuitdesigncompletion2023},
we use a one-stage YOLO detector to detect all circuit
components. Namely, the YOLO-v8 version from
Ultralytics~\cite{Jocher_Ultralytics_YOLO_2023}, which improves the
detection of small
objects~\cite{VIJAYAKUMARYOLObasedObjectDetection2024}.  Due to the
lack of a public dataset containing electrical circuits with european
symbols, the first and second semester students studying electrical
engineering at the HTWG Hochschule Konstanz drew $831$ resistor
circuits comprising linear and parallel circuits, voltage dividers,
Wheatstone bridges, and delta- and star-circuits. The selection of
circuits is based on the syllabus of electrical engineering 1. The
labeled dataset can be accessed
here:~\cite{CKnievel_AITEE_Dataset_2025}.  In addition to the passive
and active two-pole circuits, the identifiers of the two-pole circuits
as well as the corner and intersection points in the circuit have also
been labeled. Four variants of the YOLOv8 model were trained (nano,
small, medium, large) and their runtime and mean average precision
were measured at an IoU of 0.5 on a Intel i7-4790k CPU. The results
are given in Table~\ref{tab:yolo-prec}.
\begin{table}[htbp]
  \centering
  \begin{tabular}[htp]{ccc} \hline Model & Runtime in ms &
    $m$AP$_{0.5}$ \\ \hline
    YOLOv8n & 120 & $0.965$ \\
    YOLOv8s & 211 & $0.971$ \\
    YOLOv8m & 392 & $0.971$ \\
    YOLOv8l & 632 & $0.973$ \\\hline
  \end{tabular}
  \caption{Precision and runtime results for the YOLOv8-based
    detection.}
  \label{tab:yolo-prec}
\end{table}
\begin{figure}[bp]
  \centering \includegraphics[width=0.8\textwidth,
  keepaspectratio]{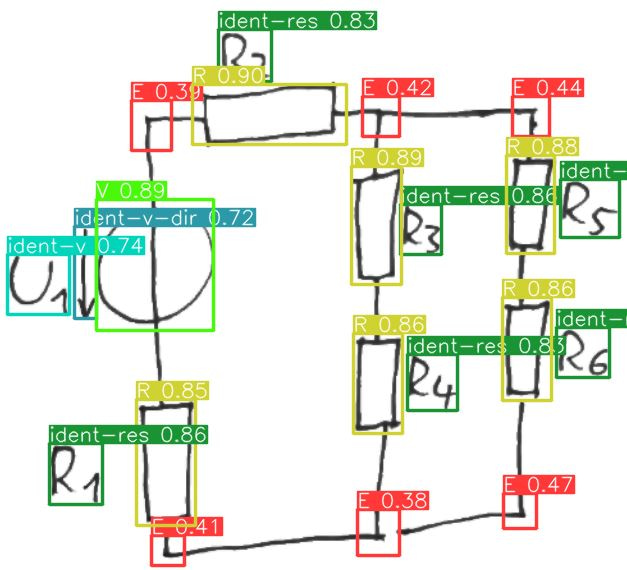}
  \caption{Output of the object detection for the example circuit with
    the YOLOv8s model.}
  \label{fig:yolo-obj-output}
\end{figure}
Based on these results, we chose the YOLOv8s model providing the best
trade-off between precision and runtime.  The output of the object
detection is shown for the example circuit in
Fig.~\ref{fig:yolo-obj-output}. Given the detection results, we can
subsequently proceed to reconstruct the connections between the
detected components.  In contrast to previous publications, we also
detect the corner and intersection points in a circuit. This
facilitates a simple approach to also detect diagonal connections. The
process of the connection recognition is shown in
Fig.~\ref{fig:edge-detect-flow}.  In a first step, all detected
components and their identifiers are removed from the image. Then,
$N_d$ contour points are created on the remaining connections (see
\circled{$c_1$} in Fig.~\ref{fig:edge-detect-flow}). In parallel, all
corner and intersection points are connected to each other building
so-called inter-node connections (see \circled{$c_2$} in
Fig.~\ref{fig:edge-detect-flow}).

The validation of inter-node connections is performed using a
line-loss metric that quantifies the geometric proximity between
candidate connections and actual circuit paths. The line-loss
computation consists of two steps: Initially, each inter-node
connection $k$ is discretized with $N_{b,k}$ equidistant interval
points $(x_{k,i}, y_{k,i})$. Subsequently, the Euclidean distance is
calculated from each interval point to its nearest circuit contour
point. The set of contour points is defined as:
\begin{align}
  \label{eq:4}
  N_d = \{(x_j, y_j) \in \mathbb{R}^2 \mid j = 1,\ldots,n\}.
\end{align}
The line-loss metric for connection $k$ is computed as:
\begin{align}
  \label{eq:line-loss}
  d_k = \sum_{i=1}^{N^-_{b,k}} \min_{c_j \in N_d}\sqrt{(x_{k,i} - x_j)^2 + (y_{k,i} - y_j)^2}
\end{align}
where $(x_{k,i}, y_{k,i})$ denotes the coordinates of the $i$-th
interval point of connection $k$, for $i \in {1,...,N_{b,k}}$. The
term $N^-_{b,k} \leq N_{b,k}$ accounts for the exclusion of interval
points which are located within a bounding box of a detected component
from the line-loss metric. The validation step establishes a
heuristically determined linear threshold value to differentiate valid
from invalid inter-node connections. The result of the analysis is
shown next to \circled{$c_3$} in Fig.~\ref{fig:edge-detect-flow} where
green lines represent the valid inter-node connections and red lines
belong to invalid inter-node connections.  The final integration step,
indicated by \circled{$c_4$} in Fig.~\ref{fig:edge-detect-flow},
compares the bounding boxes of the detected components with the valid
inter-node connections. The resulting intersections are used to
incorporate the components into the electrical circuit structure.

Additionally, each valid inter-node connection corresponds to a
netlist node. Together with the class of the detected component, this
allows both the netlist to be generated and graph-based processing to
be enabled.
\subsection{Graph Embedding \& Similarity Measure}
\label{sec:graph-embedding-}

\begin{figure}[bp]
  \centering \includegraphics[width=0.9\textwidth,
  keepaspectratio]{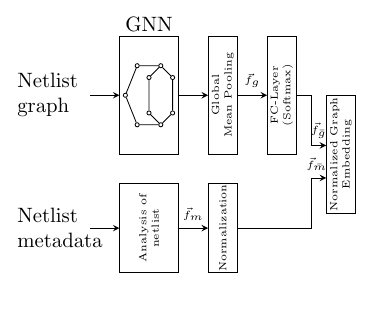}
  \caption{Process to calculate normalized graph embeddings using the
    netlist graph as well as netlist metadata.}
  \label{fig:graph-embeddings-proces}
\end{figure}

After reconstructing an electrical circuit diagram, it becomes
necessary to identify its corresponding context within the lecture
materials. Electrical engineering fundamentals are typically taught
using basic circuit configurations, including series circuits,
parallel circuits, and combinations thereof. Students face a primary
challenge in applying learned principles across different circuit
configurations. For AITEE, this presents a specific challenge since
the input circuit may not exactly match those presented in lecture
materials. Therefore, the objective is to identify the most analogous
circuit and derive the applicable methodologies. In this paper, we
propose to model the electrical circuit as an undirected graph and to
use the global graph embeddings to calculate a circuit-similarity
measure. Due to the application within an educational setting, the
similarity between electrical circuits is primarily defined by their
shared methodological approaches to problem-solving. Two key
characteristics determine the calculation methodology:
\begin{enumerate}
\item \textbf{Circuit Type}: This describes the interconnection
  pattern of components within the electrical circuit. Each circuit
  type (series, parallel, mixed, and bridge circuits) typically
  requires specific formulas and procedures for problem-solving.
\item \textbf{Special Cases}: These arise when specific conditions,
  unusual components, or particular connection types are present. Even
  a single connection or component can trigger a special case,
  potentially requiring a completely different calculation
  methodology. The superposition principle is one such special case,
  used to analyze circuits with multiple independent sources by
  evaluating each source's effect individually before combining the
  results.
\end{enumerate}
This definition of circuit similarity forms the foundation for
developing feature representations that can effectively capture these
characteristics for comparison purposes.  In order to develop an
effective feature representation, we formulate a classification
problem with eight distinct circuit classes. These classes are derived
from combining four basic circuit types (parallel, series, mixed, and
bridge circuits) with two source configurations (single and multiple
sources). The circuit classifications are summarized in
Table~\ref{tab:circuit-classes}.
\begin{table}[htbp]
  \centering
  \begin{tabular}[tbp]{l | c | c} \toprule Circuit Class & Single
    source & Multiple sources \\ \midrule
    Parallel  Circuit & Class 1 & Class 2 \\
    Series Circuit & Class 3 & Class 4 \\
    Mixed Circuit & Class 5 & Class 6 \\
    Bridge Circuit & Class 7 & Class 8\\ \bottomrule
  \end{tabular}
  \caption{The different circuit classes in the GNN classification.}
  \label{tab:circuit-classes}
\end{table}
The computation of graph embeddings follows the process illustrated in
Fig.~\ref{fig:graph-embeddings-proces} and is explained in detail in
Sec.~\ref{sec:graph-embedding}. In the evaluation of suitable
architectures for the graph neural network (GNN) component, several
established approaches were examined: Graph Convolutional Networks
(GCNs)~\cite{KIPFSemiSupervisedClassificationGraph2017}, Graph
Attention Networks (GATs)~\cite{VELICKOVICGraphAttentionNetworks2018},
GraphSAGE~\cite{HAMILTONInductiveRepresentationLearning2018}, and
Graph Isomorphism Network (GIN)~\cite{XUHowPowerfulare2019}.
The evaluation involved training the different GNNs with 150 netlists from various classes and validating them using 30 netlists. Based on this evaluation,
GraphSAGE was chosen showing slightly better performance.

\subsubsection{Graph Embedding}
\label{sec:graph-embedding}

The graph embedding generation integrates two primary inputs: the
netlist graph and its associated metadata, processed through distinct
pathways as depicted in Fig.~\ref{fig:graph-embeddings-proces}. The
netlist graph encodes component interconnections and their topological
relationships, whereas the metadata comprises the amount and type of
components. The structural information initializes the GraphSAGE
network, generating node embeddings that incorporate both local and
global structural characteristics. These node embeddings capture
contextual information from their neighborhood, yet they do not
inherently provide a comprehensive representation of the entire graph
structure. To address this limitation, a global pooling operation is
implemented, aggregating the node embeddings into a single
representative vector. A subsequent fully-connected layer with a
softmax activation function outputs the normalized feature vector
$\vec{f}_{\bar{g}}$.

The secondary path implements a heuristic approach to process netlist
metadata. This approach comprises three key components: a sigmoid
function $f_c$ for component quantification, a linear combination
$f_s$ for source type distribution and a binary function $f_b$ that
differentiates between single-source ($f_b=0$) and multi-source
($f_b=1$) configurations.

The component quantification function $f_c$ uses a sigmoid form
defined as
\begin{align}
  \label{eq:basic-sigmoid-func}
  f_{c} = \frac{1}{1+\exp\left(-c_1 \cdot (x-c_2)\right)}.
\end{align}
The sigmoid parameters were calibrated with $c_1{=}1$ and $c_2{=}7.5$,
establishing a normalized range of $[0,1]$ for circuits containing 1
to 14 components. This range covers the typical complexity found in
lecture materials.  A linear combination quantifies the number and
type of sources:
\begin{align}
  \label{eq:5}
  f_s = 0.33 \cdot V + 0.66 \cdot C + 0.01\cdot V\cdot C,
\end{align}
where $V$ and $C$ are binary indicators ($V,C \in \lbrace 0,1\rbrace$)
for the presence of voltage and current sources, respectively.

Finally, a binary function $f_b$ indicates whether there is only one
source ($f_b {=} 0$) or multiple sources ($f_b {=} 1$) in the circuit.
The feature representations of all three functions are consolidated
into a unified vector $\vec{f}_m = [\vec{f}_c, \vec{f}_s,
\vec{f}_b]$. To ensure consistent scaling, the elements of $\vec{f}_m$
are normalized, constraining their sum to unity, and stored in
$\vec{f}_{\bar{m}}$.

\subsubsection{Similarity Measure}
\label{sec:similarity-measure}
\begin{figure}[htbp]
  \centering \includegraphics[width=1\textwidth,
  keepaspectratio]{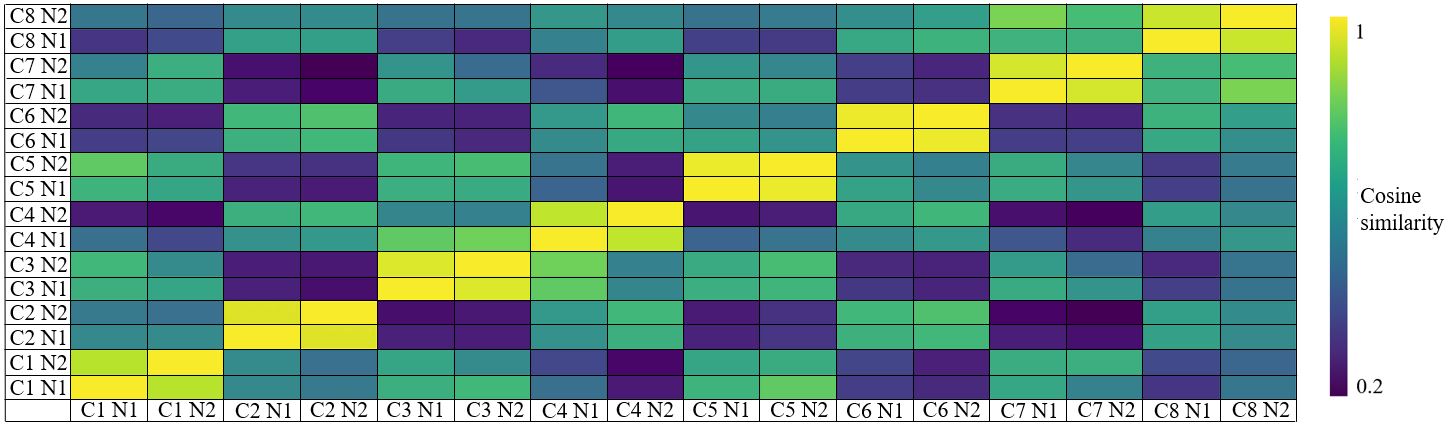}
  \caption{Cosine Similarity Map of Circuit Embeddings. Heat map
    showing similarities between circuit embeddings across 8 classes
    (2 circuits per class). High similarity values (yellow) appear
    between circuits of the same class, with minimal similarity
    (black) between different classes, demonstrating the effectiveness
    of graph embeddings in distinguishing circuit topologies.}
  \label{fig:sim-measure-graph}
\end{figure}

The effectiveness of graph embeddings for circuit representation is
clearly demonstrated in our experimental results. As shown in
Fig.~\ref{fig:sim-measure-graph}, the similarity map based on cosine
distances between embeddings across 8 distinct circuit classes (2
circuits per class) reveals strong intra-class relationships. Circuits
belonging to the same class exhibit high similarity values,
approaching 1, indicating their embeddings point in nearly identical
directions within the latent space. Conversely, cross-class
comparisons show minimal similarity, suggesting the embeddings become
increasingly orthogonal as circuit differences grow. This clear
separation validates that the graph-based representation successfully
captures the fundamental characteristics that define circuit classes
while distinguishing between different topological configurations.

\section{LLM-based Tutor in Electrical Engineering}
\label{sec:llm-based-tutor}

To ensure the technical accuracy of AITEE, it is essential that the
employed LLM is able to correctly interpret a given electrical circuit
as well as to apply corresponding solution methods. The correct
recognition and interpretation of the electrical circuit represented
by a netlist is therefore crucial. Misinterpretation at this stage can
introduce significant errors, potentially compromising the
effectiveness of domain-specific electrical engineering knowledge when
applied to an inaccurately understood circuit. The following section
analyzes the fundamental capabilities of three open-source and one
closed-source LLM in interpreting netlist representations. Subsequent
chapters will then evaluate the application of Retrieval-Augmented
Generation (RAG) approaches for solving electrical circuit
tasks. Finally, the robustness of the agent and the effectiveness of
Socratic dialogue strategies will be assessed.

\subsection{Understanding of Electrical Circuits}
\label{sec:underst-electr-circ}

For the evaluation of the LLMs' capabilities to understand electrical
circuits, we manually created a dataset comprising 24 netlists, with three examples each for the circuit classes defined in Table~\ref{tab:circuit-classes}, along with  their corresponding accurate descriptions. Each model received netlists from the dataset and was tasked
with generating circuit descriptions. The initial assessment focused
on the baseline performance of LLMs without optimization. To automate
the evaluation process, GPT-4.0 was employed as the judge,
utilizing the LLM-as-a-Judge method described by Zheng et
al~\cite{ZHENGJudgingLLMasaJudgeMTBench2023a}.  For each generated
description, the judge was instructed to provide a rating from 0 to 4,
based on the following scoring scheme:
\begin{itemize}
\item 0 points: The description is completely incorrect.
\item 1 point: The description exhibits numerous errors or fails to
  capture many aspects of the reference description.
\item 2 points: The description includes a limited number of errors or
  differs from the reference in a few aspects.
\item 3 points: The description displays only minor errors or diverges
  in a few aspects from the reference description.
\item 4 points: The description is entirely error-free and logically
  describes the same circuit as the reference description.
\end{itemize}
A corresponding prompt example for the baseline approach is shown in
Fig.~\ref{fig:baseline-prompt-judge}
\begin{figure}[htbp]
  \centering
  \begin{tcolorbox}[colback=white, colframe=black, arc=2mm,
    boxrule=1pt]
    \small{\textbf{\texttt{[Human]}}}\\
    \small{\texttt{\#\#\# Here is the netlist to be described. Netlist:}} \\
    \small{\texttt{R\_D  N004  N005  450$\Omega$}}\\
    \small{\texttt{R\_B  N002  N003  270$\Omega$}}\\
    \small{\texttt{R\_C  N003  N004  360$\Omega$}}\\
    \small{\texttt{V7\;\;  N001  N005  18V}}\\
    \small{\texttt{R\_A  N001  N002  180$\Omega$}}\\
    \small{\textbf{\texttt{[System]}}}\\
    {\texttt{Your task is to analyze a netlist and briefly and
        concisely describe the circuit. Describe the circuit
        represented by the netlist, not the netlist itself.}}
  \end{tcolorbox}
  \caption{Baseline prompt example for the generation of circuit
    descriptions for a given netlist.}
  \label{fig:baseline-prompt-judge}
\end{figure}
The baseline accuracy results are presented in the first column of
Table~\ref{tab:understanding-circuits}. Accuracy is quantified as the
ratio of the total points achieved to the maximum possible total
points.

The smallest model, Llama~3.1~8B, demonstrated a significant deficit
in netlist comprehension, which resulted in the misinterpretation of
the majority of circuits within the dataset. The next larger
open-source models, Llama~3.1~70B and Llama~3.1~405B, also showed
fundamental shortcomings in this area. A particular notable weakness
was observed in the interpretation of electrical nodes. The
closed-source model Claude~3.5~Sonnet accurately described simple
circuits such as series and parallel configurations. However, it
demonstrated limitations with more complex circuits, particularly in
the recognition of nodes and parallel branches.
\begin{table}[htbp]
  \centering
  \begin{tabular}{l|ccccc} \toprule
    \multirow{2}{*}{Model} & \multicolumn{5}{c}{Accuracy} \\
                           & \rotatebox{90}{\textit{Baseline}} &
                                                                 \rotatebox{90}{\textit{CoT}}
                                                               &
                                                                 \rotatebox{90}{\textit{2-Shot-CoT}}
                                                               &
                                                                 \rotatebox{90}{\textit{4-Shot-CoT}}
                                                               &
                                                                 \rotatebox{90}{\textit{\parbox{6em}{4-Shot-CoT
                                                                 +\\
    Contextualization}}} \\ \midrule
    Llama 3.1 8B & 0.25 & 0.28 & 0.31 & 0.5 & 0.57 \\
    Llama 3.1 70B & 0.37 & 0.69 & 0.83 & 0.87 & 0.89 \\
    Llama 3.1 405B & 0.55 & 0.73 & 0.82  & 0.89 & 0.90 \\
    Claude 3.5 Sonnet & 0.74 & 0.8 & 0.95 & 0.95 & 0.97 \\ \bottomrule
  \end{tabular}
  \caption{Accuracy for the correct interpretation and analysis of
    electrical circuits as a function of prompt engineering method by
    LLM-as-a-Judge.}
  \label{tab:understanding-circuits}
\end{table}

Chain-of-Thought (CoT)
prompting~\cite{WEIChainofThoughtPromptingElicits2023a} was
implemented to enhance reasoning capabilities of the models in the
analysis, recognition, and interpretation of netlists, which is
inherently a complex reasoning task. The previously employed baseline
prompt provided only a brief task description, prompting the LLMs to
attempt a single-step solution. To address this, the prompt was
modified to guide the LLMs to process the task through a defined chain
of thought. Specifically, the chain begins with identifying the
component connections, followed by analyzing the current flow pattern
through the circuit. The analysis then proceeds to identify circuit
topologies and configurations, examining parts of the circuit which
are in a series or parallel arrangement, delta/wye connections, or
bridge circuits. Only after completing this systematic examination
does the process generate a comprehensive circuit description. It can
be seen from the results in the second column of
Table~\ref{tab:understanding-circuits} that the Llama models 70B and
405B improve significantly while Claude Sonnet~3.5 and especially
Llama~3.1~8B only slightly improve.
\begin{figure}[htbp]
  \centering
  \begin{tcolorbox}[colback=white, colframe=black, arc=2mm,
    boxrule=1pt]
    \textbf{\texttt{[Human]}}\\
    \small{\texttt{\#\#\# Here is the netlist to be described. Netlist:}} \\
    \small{\texttt{R\_D  N004  N005  450$\Omega$}}\\
    \small{\texttt{R\_B  N002  N003  270$\Omega$}}\\
    \small{\texttt{R\_C  N003  N004  360$\Omega$}}\\
    \small{\texttt{V7\;\;    N001  N005  18V}}\\
    \small{\texttt{R\_A  N001  N002  180$\Omega$}}\\
    \small\textbf{\texttt{[System]}}\\
    \small{\texttt{Your task is to analyze a netlist and briefly and concisely describe the circuit it represents.  Follow these steps in order:}} \\
    \small{\texttt{1. Create a description explaining how the
        components of the
        circuit are connected. \\
        2. Create a description of how the electric current flows
        through
        the circuit from the first pole of the source to the second.\\
        (This point can be ignored for circuits with multiple sources) \\
        3. Create a list of sub-circuits, such as series circuits,
        parallel circuits, or delta/star connections. \\
        4. Create a description of the overall circuit. (Describe the
        circuit represented by the netlist, not the netlist itself.)}}
  \end{tcolorbox}
  \caption{Chain-of-thought prompt example for the generation of
    circuit descriptions for a given netlist.}
  \label{fig:cot-prompt-example}
\end{figure}

In order to further enhance the performance, few-shot prompting, as
described by Brown et al. \cite{BROWNLanguageModelsare2020}, was
evaluated. This technique was implemented with both two and four
examples, in conjunction with Chain-of-Thought prompting. These
configurations are denoted as 2-Shot-CoT and 4-Shot-CoT, respectively,
in Table~\ref{tab:understanding-circuits}. As can be seen from the
results, further improvements were achieved for all models. Notable
Claude~Sonnet~3.5 reached a near optimal results of $0.95$.

Building upon the initial analysis of netlist interpretations, which
revealed frequent inaccuracies in the identification of electrical
nodes, a static contextualization strategy was introduced. This approach
incorporates deterministically derived information about the
electrical nodes directly into the prompt. Furthermore, guidance on
interpreting the netlist structure was also provided within the
prompt. The contextualization in combination with 4-Shot-CoT Prompting
achieved the best results. It can be seen, that the mid-sized Llama
model (70B) achieved almost the same results as the 405B model and
performs only slightly worse than the Claude~3.5~Sonnet model.

It is important to note that although a perfect score was not achieved
by any model, the scoring was influenced by GPT-4.0 as the judge,
which lowered scores for minor deviations from the reference
description. With the exception of Llama~3.1~8B, all models are able
to provide sufficiently accurate descriptions of the netlist.

\begin{table*}[tp]
  \centering
  \begin{tabular}{l|ccccccc} \toprule
    \multirow{2}{*}{Model} & \multicolumn{7}{c}{Accuracy} \\
                           & \rotatebox{90}{\textit{Baseline}} &
                                                                 \rotatebox{90}{\textit{3-Shot-CoT}}
                                                               &
                                                                 \rotatebox{90}{\parbox{6em}{\textit{3-Shot-CoT
                                                                 {+} \\
    Naive-RAG}}} & \rotatebox{90}{\parbox{6em}{\textit{3-Shot-CoT {+}
                   RAPTOR {+} \\ RAG-Fusion}}} &
                                                 \rotatebox{90}{\parbox{6em}{\textit{3-Shot-CoT
                                                 {+} RAPTOR {+} \\
    HyDE}}}& \rotatebox{90}{\parbox{6em}{\textit{1-Shot-CoT {+} MRI}}}
                           &
                             \rotatebox{90}{\parbox{6em}{\textit{1-Shot-CoT
                             {+} MRI {+} Sim}}} \\ \midrule
    Llama 3.1 8B & 0.15 & 0.15 & 0.15 & 0.27 & 0.42 & 0.39 & 0.42\\
    Llama 3.1 70B & 0.50 & 0.57 & 0.38 & 0.65 & 0.54 & 0.77 & 0.85 \\
    Llama 3.1 405B & 0.47 & 0.68 & 0.5  & 0.65 & 0.62 & 0.85 & 0.92 \\
    Claude 3.5 Sonnet & 0.69 & 0.77 & 0.73 & 0.85 & 0.84 & 0.96 &
                                                                  0.96\\
    \bottomrule
  \end{tabular}
  \caption{Accuracy of the LLMs when applying domain-specific
    knowledge of electrical engineering to electrical circuits.}
  \label{tab:applying-methods-circuits}
\end{table*}

\begin{figure*}[tbp!]
  \centering \includegraphics{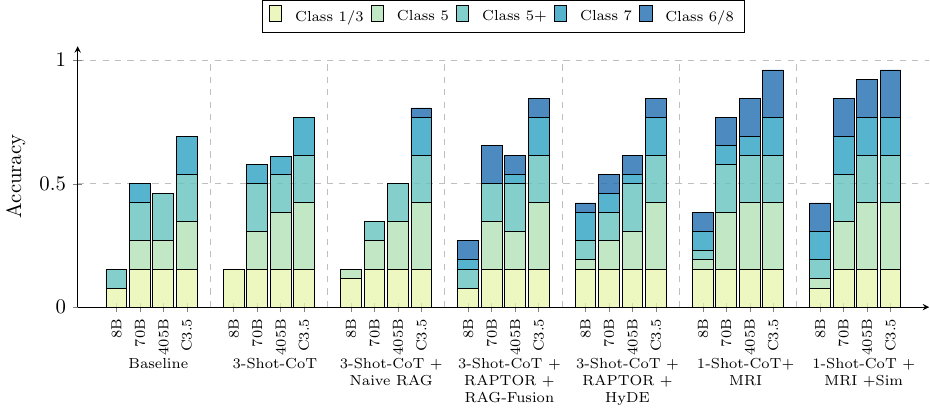}
  \caption{Accuracy by Circuit Class for the given LLM configurations. Stacked bar histogram detailing the accuracy (y-axis) of Llama 3.1 (8B, 70B, 405B) and Claude 3.5 Sonnet models under various problem-solving strategies.}
  \label{fig:acc-circ-class}
\end{figure*}

\subsection{Application of Solution Methods to Electrical Circuits}
\label{sec:appl-meth-electr}
In the following, the correct application of solution methods to tasks
for given electrical circuits is evaluated. The tasks are limited to
the curriculum of the first semester of Fundamentals of Electrical
Engineering, in which, among other topics, resistance networks with
direct current are examined. One or two tasks for a subset of circuit
classes from Table~\ref{tab:circuit-classes}, with several subtasks,
are evaluated.  In order to make a precise statement about the
capabilities of the models in relation to the correct application of
the methods, the reference description of the netlist is provided for
each task.  Since none of the models examined, including GPT-4.0,
was able to solve the tasks without errors, the solutions of all
models were checked manually. The achievable partial points were
defined in advance for each subtask to ensure a consistent evaluation.

\subsubsection{Baseline Performance Evaluation}
\label{sec:basel-perf-eval}

The baseline results over all circuit classes are shown in the first
column of Table~\ref{tab:applying-methods-circuits}. Furthermore, the
results per circuit class are depicted as a stacked bar plot in
Fig.~\ref{fig:acc-circ-class}. Due to their particular importance and
widespread use in lecture materials, we have listed tasks on voltage
and current dividers for mixed circuits separately, denoted by class
5+. As Fig.~\ref{fig:acc-circ-class} illustrates for the Baseline configurations, correct solutions are predominantly concentrated in the simpler Class 1/3 circuits (single source source, series/parallel) and, to a lesser extent, Class 5+ (voltage/current dividers). This latter observation supports the notion that tasks on this subclass could be solved significantly better due to their widespread use in training material.
The Llama 3.1 70B and 405B models were able to solve many
tasks for the simple series and parallel circuits (Class 1/3) and a majority of tasks related to current and voltage dividers (Class 5+). However, more complex configurations such as Class 7 and especially Class 6/8 saw minimal to no success across all models at baseline.

Furthermore, the baseline performance of the models could not be
significantly improved by CoT prompt engineering either. Hereby, the
number and order of the examples have been empirically evaluated and
set to three examples. The results for the 3-Shot-CoT can be seen in
the second column of
Table~\ref{tab:understanding-circuits} and detailed in Fig.~\ref{fig:acc-circ-class}. Figure~\ref{fig:acc-circ-class} confirms that 3-Shot-CoT offered only marginal gains over the baseline approach for most models, with performance still heavily reliant on solving Class 1/3 and Class 5+ circuits. While the 
Llama~3.1~405B model was able to achieve a more noticeable improvement in
performance, Fig.~\ref{fig:acc-circ-class} reveals this was largely due to an increased proficiency on these same less complex classes, rather than a breakthrough in handling more difficult circuit types.

\subsubsection{Retrieval-Augmented Generation and its Limitations}
\label{sec:retr-augm-gener}

It is evident that neither the baseline performance nor the
performance achieved with the 3-Shot-CoT approach for the language
models is sufficient for a tutor application, particularly given their challenges with circuits beyond moderate complexity.  A typical solution to
provide the domain-specific knowledge to the LLM is given by
Retrieval-Augmented Generation
(RAG)~\cite{SLADETransformingLearningAssessing2024,MULLINSEnhancingclassroomteaching2024,DONGHowBuildAdaptive2025}.
For AITEE, the lecture content was preprocessed as a knowledge base
where relevant formulas and calculations were reproduced with \LaTeX\
equations. Circuit illustrations were converted to netlists and placed
at appropriate locations. The script was then divided into 400-token
chunks. OpenAI's text-embedding-ada-002-v2 model was used to create
the embeddings. To identify semantically relevant content in the
vector database, a circuit description must be added to the prompt
alongside the student's question (e.g., "How do I calculate the
current $I_3$"). The three most similar chunks are returned and used
to contextualize the LLM. The results are denoted by 3-Shot-Cot +
Naive RAG. Compared to isolated prompt engineering, the performance
actually deteriorated for some models. A detailed analysis of the responses revealed
that the naive RAG approach introduced an additional source of
error. Without RAG, the baseline models relied on their trained
knowledge, whereas with RAG, they used the provided chunks for finding
solutions. Unsuitable chunks led to poorer responses. However,
identifying the relevant chunks is challenging. Simply combining the
circuit description and the task formulation is not sufficient to find
appropriate chunks. The query must be optimized for the retrieval
process. Additionally, some queries relate to multiple sections of the
script. For example, when a question about a mixed circuit is posed
and this circuit is simplified during the response process, such as to
a series circuit, it would be optimal to have chunks with higher
abstraction that contain information about both mixed circuits and
series circuits.

To address these limitations, we evaluated two advanced retrieval
approaches. The first approach combines
RAPTOR~\cite{SARTHIRAPTORRecursiveAbstractive2023} with
RAG-Fusion~\cite{RACKAUCKASRagFusionNewTake2024}. RAPTOR (Recursive
Abstractive Processing for Tree-Organized Retrieval) constructs a
hierarchical tree of recursively embedded, clustered, and summarized
text chunks, enabling retrieval at different levels of
abstraction. RAG-Fusion complements this by generating multiple
contextual queries and reranking them using reciprocal rank fusion,
which helps capture various perspectives of the original query.

The second approach pairs RAPTOR with HyDE (Hypothetical Document
Embeddings)~\cite{GAOPreciseZeroShotDense2023}.  which specifically
addresses the style mismatch between student queries and the knowledge
base. HyDE first uses a large language model to generate a
hypothetical text segment that mimics the style of the lecture script while answering the query. Although this generated text
may contain errors, it is not used directly for answering but rather
to identify semantically similar chunks in the vector database. This
approach is particularly valuable in educational contexts where
first-semester students' questions often differ significantly from the
formal language used in lecture scripts. By bridging this linguistic
gap, HyDE enables more effective retrieval of relevant information
despite differences in formulation and terminology.

Although advanced RAG methods improved the performance of the naive
RAG approach, as illustrated in Fig.~\ref{fig:acc-circ-class}, they enabled models to solve a greater proportion of Class 5, Class 5+, and begin addressing Class 7 circuits. However, for the Llama models, these methods did not achieve significantly better performance compared to isolated prompt engineering (cf. 3-Shot-CoT performance), especially for the most complex classes.
A primary limitation was the suboptimal
suitability of queries - comprising circuit descriptions and questions
- for similarity searches of matching chunks. Only the large
closed-source Claude 3.5 Sonnet model achieved a sufficient overall
performance and, as seen in Fig.~\ref{fig:acc-circ-class}, a broader capability across circuit complexities  with advanced RAG methods to suggest tutor-level
expertise.

\subsubsection{Multi-Representation Indexing for Improved Retrieval}
\label{sec:multi-repr-index}

A key consideration for RAG is the chunking strategy. Instead of
segmenting content based on a fixed number of tokens, the teaching
material is structured into clearly defined units. From a didactic
perspective, a unit represents basic building blocks of knowledge in
electrical engineering, encompassing declarative knowledge
(definitions, facts), procedural knowledge (application methods,
problem-solving strategies), and conceptual knowledge (understanding
of interrelationships and principles). Thus, supplementing the prompt
with the most relevant unit is anticipated to enhance the performance
of the LLM.

To address the query suitability issue, multi-representation indexing
(MRI) was implemented. Chen et al.~\cite{CHENDenseRetrievalWhat2024}
introduced MRI, advocating for indexing a corpus using
``propositions'' – concise, self-contained factoids – as retrieval
units. In contrast to this proposition-based approach, our
implementation of MRI utilizes representative netlists as indices for
units. For each unit, typical electrical circuits are generated, and
their corresponding netlists serve as indices for that unit. When a
prompt includes a circuit, the GNN-based similarity measure, detailed
in Section~\ref{sec:graph-embedding-}, identifies the representative
netlists most similar to the given circuit. The units associated with
these similar netlists are then retrieved and provided to the LLM in
combination with a single CoT example. The performance results of this
approach, termed \textit{1-Shot-CoT${+}$MRI}, are presented in
Table~\ref{tab:applying-methods-circuits} and
Figure~\ref{fig:acc-circ-class}.

The implemented system demonstrates a significant performance
improvement compared to previously evaluated approaches. As shown in Fig.~\ref{fig:acc-circ-class}, the 1-Shot-CoT${+}$MRI approach led to a substantial increase in accuracy, particularly enabling models to successfully address more complex circuit classes. With the
exception of the Llama~3.1~8B model, all other models exhibit a
performance level that suggests the potential to ensure tutor-level
expertise. Notably, Llama~3.1~70B and 405B, and especially the Claude~3.5~Sonnet model, showed significant capability in solving Class 7 and even the challenging Class6/8 problems, which were largely unsolvable with previous methods. The Claude~3.5~Sonnet model achieves near-flawless
performance on the tasks within the dataset.

Consistent with findings reported by Chen et
al.~\cite{CHENImpactLanguageArithmetic2024}, our analysis also reveals
a recurring challenge for all language models in performing basic
arithmetic operations.

\subsubsection{Simulation-Based Arithmetic Validation}
\label{sec:simul-based-arithm}

To address the identified limitations of LLMs in arithmetic
operations, the system was augmented with a simulation execution
capability. This enhancement incorporates the tool
PySpice~\cite{PySpice}. The netlist representation of the circuit is
provided as input to PySpice, and the simulation generates output
parameters including partial voltages, currents, total current, total
voltage, and total resistance.

The results presented in Table~\ref{tab:applying-methods-circuits} and
Figure~\ref{fig:acc-circ-class} with 1-Shot-CoT${+}$MRI${+}$Sim,
indicate near-optimal performance for both the Llama~3.1~405B and
Claude~3.5~Sonnet models across most circuit classes. However, for
tasks within Classes 6 and 8, which necessitate the application of the
superposition principle, some inaccuracies in current calculations
were observed. These errors appear to originate from inconsistencies
in current direction definitions between the provided netlist and the
task query. Specifically, the system may have failed to detect or
reconcile cases where the netlist's current direction convention
deviated from that implied or explicitly stated in the query,
resulting in incorrect calculations using the superposition method.

\subsection{Evaluation of Didactic Competence}
\label{sec:eval-didact-comp}

A main goal of AITEE is to generate didactically valuable
responses. This necessitates that the tutor guides students towards
solutions, rather than directly presenting them.  While a
comprehensive analysis of the full spectrum of didactic capabilities
in LLMs presents a significant challenge, this section concentrates on
evaluating key aspects of pedagogical effectiveness relevant to a
tutoring system. Specifically, we focus on two critical dimensions of
didactic quality: fostering learner autonomy and dialogue
robustness. These two metrics are prioritized as essential indicators
of a system's ability to provide effective and pedagogically sound
guidance.  To provide a focused and evaluable assessment of didactic
quality, we employ the following metrics:

\textbf{Fostering Learner Autonomy}: This metric assesses the system's
success in promoting independent learning. Recognizing that effective
tutoring should guide rather than dictate, we evaluate whether the
system avoids directly providing solutions or explicit intermediate
steps. Instead, pedagogically sound dialogues are expected to employ
counter-questions and guiding prompts to facilitate the learner's
autonomous progress towards both intermediate and final
solutions. Dialogues are considered to fall short in fostering
autonomy if the system preempts the learner's problem-solving process
by directly supplying final answers or critical intermediate results.

\textbf{Dialogue Robustness}: This metric specifically measures the
system's resilience to potentially inaccurate user input. A key
characteristic of a robust tutoring agent is its ability to maintain a
consistent and correct understanding, even when confronted with
erroneous user statements. For example, a robust system should remain
unaffected if a user mistakenly classifies a series circuit as a
parallel circuit. To specifically examine dialogue robustness, each
evaluation dialogue includes a simulated instance of such
user-provided misinformation. Dialogues are classified as
insufficiently robust if the system inappropriately accepts the
inaccurate user statement and subsequently adapts its behavior based
on this error.

In order to ensure a focused evaluation, a dataset was constructed
consisting of electrical circuit descriptions paired with
corresponding tasks or questions. Each question–circuit pair serves as
the starting point for a dialogue, which is then extended to include
five user queries and five system responses. To assess dialog
robustness, each conversation includes one intentional insertion of
false information (for example, incorrectly labeling a parallel
circuit as a series circuit). This methodology results in five
dialogs, each containing five question–answer exchanges per initial
query. Finally, the resulting dialogs were evaluated using the
predefined metrics for learner autonomy and dialogue robustness. The
results are presented in Table~\ref{tab:robust-llms}.

All evaluated models exhibit fundamental behavioral deficiencies in
the context of this tutoring application. Specifically, the LLMs
consistently generated complete solutions directly, a practice that
could negatively impact student learning outcomes. Regarding dialogue
robustness, the smallest model, Llama~3.1~8B, adopted the user's
perspective in four out of five dialogues. This behavior was also
observed in the other models, albeit less frequently, occurring twice
out of five dialogues. In all cases, this level of robustness is
deemed insufficient for effective pedagogical application.
\begin{table}[htbp]
  \centering
  \begin{tabular}{l|cc|cc} \toprule
    \multirow{2}{*}{Model} & \multicolumn{2}{c}{\parbox{6em}{\centering Learner\\ Autonomy}} &  \multicolumn{2}{c}{\parbox{6em}{\centering Dialogue Robustness}} \\
                           & \rotatebox{90}{\textit{Baseline}} &
                                                                 \rotatebox{90}{\textit{w. Instructions}}
                           & \rotatebox{90}{\textit{Baseline}} &
                                                                 \rotatebox{90}{\textit{w. Instructions}}
    \\ \midrule
    Llama 3.1 8B & 0/5 & 4/5 & 1/5 & 4/5\\
    Llama 3.1 70B & 0/5 & 5/5 & 3/5 & 5/5\\
    Llama 3.1 405B & 0/5 & 5/5 & 3/5 & 5/5\\
    Claude 3.5 Sonnet & 0/5 & 5/5 & 3/5 & 5/5 \\ \bottomrule
  \end{tabular}
  \caption{Evaluation of fostering the learner autonomy and dialogue
    robustness for baselines models vs models with instruction
    prompts.}
  \label{tab:robust-llms}
\end{table}

To address these limitations, the system prompt of the LLMs is
designed to clearly define the tutors tasks and provide specific
guidelines to follow:
\begin{enumerate}
\item \textbf{Socratic Questioning}: Ask a specific question that
  stimulates the students' critical thinking and lead them step by
  step to the solution.
\item \textbf{No direct solutions}: Never provide complete or partial
  solutions. Your role is to enable students to solve problems
  independently.
\item \textbf{Promote self-efficacy}: Encourage students to think for
  themselves and apply their knowledge. Don't show the students how to
  do it, but encourage them to find the solution themselves.
\item \textbf{Error correction}: If students give incorrect answers,
  gently guide them in the right direction without giving away the
  correct answer.
\item \textbf{Technical terms}: Use and explain relevant electrical
  engineering terms to deepen understanding.
\item \textbf{Language}: Answer in German only.
\item \textbf{Adaptability}: Adapt your explanations and questions to
  the student's level of understanding.
\item \textbf{Positive reinforcement}: Reward correct answers and
  progress to increase motivation.
\item \textbf{Short and specific answers}: Always answer the student's
  specific question to enable step-by-step problem solving.
\end{enumerate}

To further align the language model with the task, few-shot examples
of desired dialogues are provided. As a result, all examined language
models engage in Socratic dialogue. Neither the closed-source model
Claude 3.5 Sonnet nor the open-source models Llama~3.1~405B and 70B
provide complete or partial results for the test dialogues. Only the
smallest model, Llama~3.1~8B, provided partial results in one of five
dialogues.

When faced with incorrect user input, the three largest models
examined remain robust and do not adopt the student's opinion. They
guide the student through the task and provide only the necessary
support. The models appropriately decline when students request
complete solutions, explaining that providing answers directly is not
possible.

% \vspace{-0.1cm}
\section{Conclusions}
\label{sec:conclusion}
This paper introduces AITEE, an agentic tutor designed to address the
limitations of traditional educational technologies in electrical
engineering education, particularly the teacher bandwidth
problem. AITEE integrates Large Language Models within an Intelligent
Tutoring System to provide interactive and personalized learning
experiences for students analyzing electrical circuits. A key feature
of AITEE is its ability to process both digital and hand-drawn circuit
diagrams, enabling students to interact with the system using either
digital tools or hand sketches. The core strength of AITEE lies in its
agentic nature, leveraging tools such as circuit reconstruction and
Spice simulation, while separately employing Socratic dialogue as its
pedagogical approach to foster learner autonomy and self-efficacy by
guiding students towards solutions through systematic questioning
rather than providing direct answers.

Our evaluation focused on
netlist interpretation and the application of domain-specific
knowledge to engineering tasks for students. Results demonstrate that
the proposed graph-based similarity measure effectively retrieves
relevant contextual information from lecture materials. Regarding
didactic competence, initial evaluations revealed a tendency for LLMs
to provide direct solutions, which hindered learner autonomy. However,
implementing instruction prompts that explicitly guide the LLMs to
adopt Socratic questioning techniques significantly improved the
system's ability to foster learner autonomy and enhance dialogue
quality. While improving dialogue robustness remains an ongoing
challenge, the instruction-prompted models demonstrated significant
improvement in resisting inaccurate user input while maintaining
pedagogical soundness.  Despite these promising results, certain
limitations persist. Arithmetic inaccuracies, particularly in complex
circuits requiring superposition, and the need for further enhancement
of dialogue robustness are identified as key areas for future work. A
crucial next step involves conducting a comprehensive test with
students to evaluate AITEE's effectiveness in real-world educational
settings and to gather feedback on its usability and impact on student
learning outcomes.
%
% \vspace{-0.15cm}
%

\bibliographystyle{IEEEtran} \bibliography{IEEEabrv, tutor}

% \begin{IEEEbiography}
%   [{\includegraphics[width=1in, height=1.25in, clip,
%   keepaspectratio]{pics/bio_knievel.jpg}}]{Christopher Knievel}
%   received the Dipl.-Ing. in electrical engineering and information
%   technology from the University of Kiel, Germany, in 2009 and the
%   Dr.~Ing. from the Chair of Information and Coding Theory,
%   University of Kiel, in 2014, respectively.

%   From January 2014 to February 2021 he was a team leader and
%   development engineer for driver assistance systems with
%   Continental AG, Lindau, Germany. Since March 2021 he is with the
%   HTWG University of Applied Sciences, Konstanz, where he is a
%   professor for autonomous systems. His research interests include
%   situation assessment, cooperative maneuver planning as well as
%   machine learning applied for mobile robots.
% \end{IEEEbiography}

\end{document}